\documentclass[11pt,a4paper]{article}
\usepackage{jcappub}

\usepackage{bm}
\usepackage{epsfig}
\usepackage{citesort}
\usepackage{graphicx}
\usepackage{amssymb}
\usepackage{color}
\usepackage{subfigure}

\newcommand{\agt}{\mbox{\;\raisebox{.3ex}
  {$>$}$\!\!\!\!\!$\raisebox{-.9ex}{$\sim$}}\;}

\newcommand{\be}{\begin{equation}}
\newcommand{\ee}{\end{equation}}
\newcommand{\bea}{\begin{eqnarray}}
\newcommand{\eea}{\end{eqnarray}}

\begin{document}


\subheader{\hfill MPP-2011-114}

\title{Impact of eV-mass sterile neutrinos on neutrino-driven supernova outflows}

\author[a]{Irene Tamborra,}
\author[a]{Georg G. Raffelt,}
\author[b]{Lorenz H\"udepohl}
\author[b]{and Hans-Thomas Janka}

\affiliation[a]{Max-Planck-Institut f\"ur Physik
(Werner-Heisenberg-Institut)\\
F\"ohringer Ring 6, 80805 M\"unchen, Germany}

\affiliation[b]{Max-Planck-Institut f\"ur Astrophysik\\
Karl-Schwarzschild-Str.~1, 85748 Garching, Germany}

\emailAdd{tamborra@mpp.mpg.de}
\emailAdd{raffelt@mpp.mpg.de}
\emailAdd{lorenz@mpa-garching.mpg.de}
\emailAdd{thj@mpa-garching.mpg.de}

\abstract{Motivated by recent hints for sterile neutrinos from the
  reactor anomaly, we study active-sterile conversions in a
  three-flavor scenario (2 active + 1 sterile families) for three
  different representative times during the neutrino-cooling evolution
  of the proto-neutron star born in an electron-capture supernova. In
  our ``early model'' (0.5~s post bounce), the $\nu_e$-$\nu_s$ MSW
  effect driven by $\Delta m^2=2.35~{\rm eV}^2$ is dominated by
  ordinary matter and leads to a complete $\nu_e$-$\nu_s$ swap with
  little or no trace of collective flavor oscillations. In our
  ``intermediate'' (2.9~s p.b.) and ``late models'' (6.5~s p.b.),
  neutrinos themselves significantly modify the $\nu_e$-$\nu_s$ matter
  effect, and, in particular in the late model, $\nu\nu$ refraction
  strongly reduces the matter effect, largely suppressing the overall
  $\nu_e$-$\nu_s$ MSW conversion. This phenomenon has not been
  reported in previous studies of active-sterile supernova neutrino
  oscillations. We always include the feedback effect on the electron
  fraction $Y_e$ due to neutrino oscillations. In all examples, $Y_e$
  is reduced and therefore the presence of sterile neutrinos can
  affect the conditions for heavy-element formation in the supernova
  ejecta, even if probably not enabling the r-process in
    the investigated outflows of an electron-capture supernova.  The
    impact of neutrino-neutrino refraction is strong but complicated,
    leaving open the possibility that with a more complete treatment,
    or for other supernova models, active-sterile neutrino oscillations
    could generate conditions suitable for the r-process.}

\maketitle

\section{Introduction}                        \label{sec:introduction}
Sterile neutrinos are hypothetical gauge-singlet fermions that could
mix with one or more of the active states and thus show up in
active-sterile flavor oscillations. Low-mass sterile neutrinos have
been invoked to explain the excess $\bar{\nu}_e$ events in the LSND
experiment
\hbox{\cite{Aguilar:2001ty,Strumia:2002fw,GonzalezGarcia:2007ib}} as well
as the MiniBooNE excess events in both neutrino and antineutrino
channels. Interpreted in terms of flavor oscillations, the MiniBooNE
data require CP violation and thus no less than two sterile
families~\cite{AguilarArevalo:2008rc,AguilarArevalo:2009xn,Karagiorgi:2009nb}.
However, these models show tension with other oscillation data and
may require additional ingredients such as non-standard
interactions~\cite{Akhmedov:2010vy}. Moreover, part of the parameter
space has been excluded by IceCube data~\cite{Razzaque:2011ab}. On
the other hand, the cosmic microwave background
anisotropies~\cite{Reid:2009nq,GonzalezGarcia:2010un,Hamann:2010bk,Giusarma:2011ex,Hou:2011ec}
and big-bang nucleosynthesis \cite{Izotov:2010ca,Aver:2010wq}
suggest cosmic excess radiation compatible with one family of sub-eV
sterile neutrinos. However, for eV-mass sterile neutrinos to be
cosmologically viable, additional ingredients are
required~\cite{Hamann:2011ge}.

Our study is motivated by the most recent indication for the possible
existence of eV-mass sterile neutrinos coming from a new analysis of
reactor $\bar\nu_e$ spectra and their distance and energy
variation~\cite{Mention:2011rk,Huber:2011wv,Kopp:2011qd}. The data
suggest a $\nu_e$-$\nu_s$ mixing of $\sin^2 2\theta \sim 0.14$ and a
mass splitting of $\Delta m^2 \agt 1.5~{\rm eV}^2$. In the supernova
(SN) context, these parameters imply that the $\nu_e$ flux would
undergo MSW conversions to $\nu_s$ closer to the SN core than any
other oscillation effects. (We assume that, because of cosmological
neutrino mass limits, the sterile state is heavier than the active
ones so that the MSW effect would occur between $\nu_e$ and $\nu_s$
and not between $\bar\nu_e$ and $\bar\nu_s$.) Even then, however, the
conversion probably would not be close enough to the SN core to affect
shock reheating during the accretion phase. Moreover, removing the
$\nu_e$ flux would stabilize the remaining flux of active neutrinos
against collective flavor conversions which anyway are likely
irrelevant for shock
reheating~\cite{Chakraborty:2011gd,Sarikas:2011am,Dasgupta:2011jf,Suwa:2011ac}.
At a larger radius, the lost $\nu_e$ flux would be partly replenished
by active-active MSW oscillations, in detail depending on the mixing
parameters among active neutrinos. So while the $\nu_e$ flux arriving
at Earth from the next nearby SN would be significantly modified,
observational signatures would probably require a large $\nu_e$
detector, in contrast to the existing large detectors that primarily
measure the arriving $\bar\nu_e$ flux by inverse beta decay. Still,
possible observational signatures may deserve a dedicated study.

We here focus on a different aspect of $\nu_e$-$\nu_s$ oscillations
that could have an interesting impact during the neutrino-cooling
phase of the proto-neutron star (PNS). The neutrino-driven matter
outflow contributes to SN nucleosynthesis and, in particular, it is a
candidate site for the formation of elements beyond iron by the
rapid neutron-capture process (for a review, see~\cite{Arnould:2007gh}
and references therein). This
``r-process'' requires a neutron-rich environment, i.e.\ an electron
fraction per baryon $Y_e< 0.5$, sufficiently large entropy to favor
a high ratio of free neutrons to ``seed'' nuclei (the latter are
usually iron-group nuclei reassembled from free nucleons and acting
as starting points of neutron captures), and sufficiently fast
timescales to lower the efficacy of converting alpha particles to
heavier nuclei. In standard SN simulations, these conditions have
remained elusive. The idea that removing the $\nu_e$ flux by
active-sterile oscillations can favor a neutron-rich outflow
environment
is not new \cite{Beun:2006ka,Keranen:2007ga,Fetter:2002xx,Fetter:2000kf,%
McLaughlin:1999pd,Hidaka:2007se,Nunokawa:1997ct}. However, the used
mass differences were larger and the possible impact of collective
active-active oscillations~\cite{Duan:2010bg} was not taken into
account. On the other hand, several recent
studies have considered the role of collective flavor oscillations
(without sterile neutrinos) on nucleosynthesis processes like the
r-process and the $\nu$p-process in
SN outflows~\cite{Duan:2010af,MartinezPinedo:2011br}.

In our study we explore the impact of $\nu_e$-$\nu_s$ oscillations
on the electron fraction $Y_e$, based on a self-consistent SN model
and the corresponding flavor-dependent neutrino fluxes from a
spherically symmetric (1D) simulation of an exploding
electron-capture SN, which leaves behind a
neutrino-cooling PNS~\cite{Huedepohl:2009wh}. To this end we begin
in section~2 with a description of our electron capture SN reference
model, we define our notation and fix the neutrino mixing
parameters. In section~3 we describe the $Y_e$ evolution in SN outflows.
In section~4 we present our results for three representative times
after core bounce ($t=0.5$, 2.9 and 6.5~s). Conclusions and perspectives
are presented in section~5.

\section{Input for neutrino flavor evolution in electron-capture supernovae}
\label{sec:inputs}

Electron-capture supernovae, originating from low-mass progenitors
(8--$10\,M_{\odot}$), might represent up to about 30\% of all
core-collapse supernovae~\cite{Wanajo:2008bw,Poelarends:2007ip}.
We use long-term simulations of a representative progenitor with mass
$8.8\,M_{\odot}$~\cite{Huedepohl:2009wh}, performed with the equation
of state of Shen et al.~\cite{Shen:1998gq}. For the present study
we chose Model Sf~21 (see reference~\cite{Huedepohl:2009wh} for further
details; the number 21 denotes a recomputation of the published model
with 21 energy bins in the neutrino transport instead of the standard
17 bins). In the chosen model, the accretion phase ends already at
$\sim 0.2$ s post bounce when neutrino heating reverses the infall
to an explosion, and the subsequent deleptonization and cooling of the
PNS take $\sim 10$~s. In this section, we discuss
our reference model for an electron-capture SN, the neutrino fluxes,
and the flavor evolution equations.

\subsection{Reference neutrino signal from electron-capture SNe}

At a radius $r$, the unoscillated spectral number fluxes for flavor
$\nu_\beta$ ($\beta = e, \bar{e}, x$ with $x = \mu$ or~$\tau$) are
\begin{equation}
F_{\nu_\beta}(E)= \frac{L_{\nu_\beta}}{4 \pi r^2}\,\frac{f_{\nu_\beta}(E)}{\langle E_{\nu_\beta} \rangle} \ ,
\end{equation}
where $L_{\nu_\beta}$ is the luminosity for flavor $\nu_\beta$, $\langle
E_{\nu_\beta} \rangle$ the mean energy, and $f_{\nu_\beta}(E)$ a quasi-thermal
spectrum. We describe it schematically in the form~\cite{Keil:2002in}
\begin{equation}
f_{\nu_\beta}(E)=\xi_\beta \left(\frac{E}{\langle E_{\nu_\beta} \rangle}\right)^{\alpha_\beta} e^{-(\alpha_\beta+1) E/\langle E_{\nu_\beta} \rangle}\ .
\end{equation}
The parameter $\alpha_\beta$ is defined by $\langle E_{\nu_\beta}^2
\rangle/\langle E_{\nu_\beta} \rangle^2 =
(2+\alpha_\beta)/(1+\alpha_\beta)$ and $\xi_\beta$ is a normalization
factor such that $\int~dE \, f_{\nu_\beta}(E)=1$. We choose
three representative times during the PNS cooling: $t = 0.5$, 2.9 and 6.5~s
after core bounce.
In table~\ref{tab:table1}, the neutrino-sphere radius, the luminosity
$L_{\nu_\beta}$, the average energies $\langle E_{\nu_\beta} \rangle$, and the
factor $\alpha_\beta$ are reported. Concerning the neutrino emission
geometry, we
adopt the spherically-symmetric bulb model~\cite{Duan:2006an}, with
the neutrino-sphere radii assumed equal for all active flavors.

\begin{table}[t]
\begin{center}
\begin{tabular}{ccccccccccc}
\hline
$t  $ & $R_\nu$ & $L_{\nu_e}$  &  $L_{\bar{\nu}_e}$  & $L_{\nu_x}$ &
 $\langle E_{\nu_e} \rangle$  &
$\langle E_{\bar{\nu}_e} \rangle$  & $\langle E_{\nu_x} \rangle$ & $\alpha_e$ & $\alpha_{\overline e}$ & $\alpha_x$\\
\hline
0.5 & 25 & 9.5 & 10.1 & 10.8 & 16.8 & 18.1 & 18.3 & 2.9 & 3.0 & 2.8\\
2.9 & 16 & 3.3 & 3.4 & 3.7 & 15.8 & 16.3 & 15.7 & 3.1 & 2.6 & 2.5\\
6.5 & 14.5 & 1 & 0.99 & 1.04 & 12.4 & 11.9 & 11.8 &  2.6 & 2.3 & 2.4\\
\hline
\end{tabular}
\vspace{2.mm} \caption{Parameters for the three considered,
representative post-bounce times (in seconds) of our SN explosion
model.
Neutrino-sphere radius $R_\nu$ in km (assumed equal for all flavors),
flavor-dependent luminosities $L_{\nu_\beta}$ (in
$10^{51}$~erg~s$^{-1}$), average energies $\langle E_{\nu_\beta} \rangle$
(in MeV), and the spectral shape factor $\alpha_\beta$ are
listed.\label{tab:table1}}
\end{center}
\end{table}

\subsection{Neutrino mixing parameters and flavor evolution equations}

Our work is motivated by the reactor anti-neutrino anomaly that
requires, if interpreted in terms of sterile neutrinos $\nu_s$,
sizable $\nu_e$--$\nu_s$ mixing. For simplicity we will
ignore possible mixings of $\nu_s$ with other active flavors.
The mass difference would have to be in the eV range, so
cosmological hot dark matter limits imply that the sterile state
would have to be heavier than the active flavors. Besides
$\nu_e$--$\nu_s$ oscillations, we also include active-active
oscillations driven by the atmospheric mass difference between
$\nu_e$ and $\nu_x$ and the mixing angle $\Theta_{13}$. The
$\nu_\mu$ and $\nu_\tau$ fluxes in a SN are very similar and these
two flavors play symmetric roles. Therefore, it is useful to define
a linear combination that is essentially identical with the $m_3$
mass eigenstate and mixes with $\nu_e$ by means of the small $\Theta_{13}$
mixing angle, and another combination that mixes through the solar
angle $\Theta_{12}$ and is separated with the solar mass difference
$\delta m_{\rm sol}$. Oscillations driven by these parameters tend to
take place at a larger radius and likely do not affect SN
nucleosynthesis. Overall, therefore, we study a 3-flavor problem
consisting of $\nu_e$, $\nu_x$ and $\nu_s$ with the mass
splittings
\begin{eqnarray}
\delta m^2_{\rm atm} &=&-2\times 10^{-3}\mathrm{\ eV}^2\ ,\\
\delta m^2_{\rm s} &=& 2.35 \mathrm{\ eV}^2\ ,
\end{eqnarray}
where the latter is representative for the reactor-inspired
values~\cite{Mention:2011rk}. Note that we assume normal hierarchy
for the sterile mass-squared difference $\delta m_{\rm s}^2>0$ and
inverted hierarchy for the atmospheric difference, $\delta m_{\rm
atm}^2<0$ for the atmospheric sector. The associated ``high'' (H)
and ``sterile'' (S) vacuum oscillation frequencies are then
\begin{eqnarray}
\omega_{\rm H} &=& \frac{\delta m^2_{\rm atm}}{2E} = -\frac{5.07}{E(\rm{MeV})} \rm{km}^{-1}\label{omegaL}\ ,\\
\omega_{\rm S} &=& \frac{\delta m^2_{\rm s}}{2E} = \frac{5.96 \times 10^3}{E(\rm{MeV})} \rm{km}^{-1} \label{omegaH}\ ,
\end{eqnarray}
whereas the usual ``low'' (L) frequency corresponds to the solar mass
difference. For the active-sterile mixing we use
\begin{equation}
\label{theta13}
\sin^2 2 \Theta_{14}= 0.165 \ .
\end{equation}
We assume a small mixing between the active flavors,
\begin{equation}
\label{theta12}
\sin^2\Theta_{13}=10^{-4}\ .
\end{equation}
In this case the MSW effect driven by this mixing angle is
non-adiabatic, i.e.\ we only focus on active-sterile MSW oscillations
and collective active-active oscillations. Recent hints for a
not-very-small value for $\Theta_{13}$ \cite{Fogli:2011qn} would
imply that we also need to include flavor conversion by the
active-active MSW effect, but this has little impact on our results.

We treat neutrino oscillations in terms of the usual matrices of
neutrino densities $\rho_E$ for each neutrino mode with energy $E$
where diagonal elements are neutrino densities, off-diagonal
elements encode phase information caused by flavor oscillations.
Moreover, we work in the single-angle approximation where it is
assumed that all neutrinos feel the same average neutrino-neutrino
refractive effect. The radial flavor variation of the
quasi-stationary neutrino flux is given by the ``Schr\"odinger
equation''
\begin{equation}\label{eq:eom1}
\mathit{i}\partial_r\rho_E=[{\sf H}_{E},\rho_{E}]
\quad\hbox{and}\quad
\mathit{i}\partial_r\bar\rho_E=[\bar{\sf H}_{E},\bar\rho_{E}]\,,
\end{equation}
where an overbar refers to antineutrinos and sans-serif letters
denote $3{\times}3$ matrices in flavor space consisting of $\nu_e$,
$\nu_x$ and $\nu_s$. The initial conditions are
$\rho_{E}=\mathrm{diag}(n_{\nu_e},n_{\nu_x},0)$ and $\bar\rho_{E} =
\mathrm{diag}(n_{\bar{\nu}_e},n_{\bar{\nu}_x},0)$. The Hamiltonian matrix
contains vacuum, matter, and neutrino--neutrino terms
\begin{equation}
{\sf H}_{E}= {\sf H}^{\rm vac}_{E}+{\sf H}^{\rm m}_{E}+{\sf H}^{\nu\nu}_{E}\ .
\label{eq:ham}
\end{equation}
In the flavor basis, the vacuum term is a function of the mixing
angles and the mass-squared differences
\begin{equation}
{\sf H}^{\rm vac}_{E} = {\sf U}\,\mathrm{diag}\left(-\frac{\omega_{\rm H}}{2},+\frac{\omega_{\rm H}}{2},\omega_{\rm S}\right) {\sf U}^{\dagger}\ ,
\end{equation}
where ${\sf U}$ is the unitary mixing matrix transforming between
the mass and the interaction basis. The matter term includes both
charged-current (CC) and neutral-current (NC) contributions and it is in
the flavor basis spanned by $(\nu_e,\nu_x,\nu_s)$
\begin{eqnarray}
\label{lambda}
{\sf H}^{\rm m} &=& \sqrt{2}G_{\rm F}\;
{\rm diag}(N_{e}-\frac{N_{n}}{2},-\frac{N_{n}}{2},0)\;,
\end{eqnarray}
where  $N_{e}$ is the net electron number density (electrons minus
positrons), and $N_{n}$ the neutron density.

In all neutral media, $Y_e=Y_p$ and $Y_n=1-Y_e$, where $Y_j$ is the
number density of particle species $j$ relative to baryons. The
local electron fraction is
\begin{equation}
\label{yedef}
Y_e(r) =\frac{N_e(r)}{N_e(r)+N_n(r)}\ .
\end{equation}
Inserting the previous expression for $Y_e$ in
equation~(\ref{lambda}), the matter Hamiltonian becomes
\begin{eqnarray}
{\sf H}^{\rm m} &=& \sqrt{2}G_{\rm F} N_b \;
{\rm diag}\left(\frac{3}{2} Y_e -\frac{1}{2},\frac{1}{2} Y_e -\frac{1}{2},0\right)\;,
\end{eqnarray}
where $N_b$ is the baryon density. The matter potential can be
positive or negative. For $Y_e>1/3$ it is $\nu_e$ that can undergo
an active-sterile Mikheev-Smirnov-Wolfenstein (MSW) resonance,
whereas for $Y_e < 1/3$ it is $\bar{\nu}_e$ \cite{wolf}.

The corresponding 3$\times$3 matrix caused by neutrino-neutrino
interactions vanishes for all elements involving sterile
neutrinos~\cite{Sigl:1992fn}, i.e. ${\sf H}^{\nu\nu}_{es}={\sf
H}^{\nu\nu}_{xs}={\sf H}^{\nu\nu}_{ss}=0$, and only the
2$\times$2 block involving the active flavors is non-zero. In
particular, the only non-vanishing off-diagonal element of the
3$\times$3 matrix is ${\sf H}^{\nu\nu}_{ex}$.

In summary, the matter plus neutrino-neutrino part of the
Hamiltonian has the diagonal elements
\begin{eqnarray}\label{lambdaes}
{\sf H}^{{\rm m}+\nu\nu}_{ee}
&=&\sqrt{2}G_{\rm F}\left[N_b\left(\frac{3}{2} Y_e - \frac{1}{2}\right) + 2 (N_{\nu_e} - N_{\bar\nu_e}) + (N_{\nu_x} -N_{\bar\nu_x}) \right]\ ,
\\
{\sf H}^{{\rm m}+\nu\nu}_{xx}
&=&\sqrt{2}G_{\rm F}\left[N_b\left(\frac{1}{2} Y_e - \frac{1}{2}\right) + (N_{\nu_e} - N_{\bar\nu_e}) + 2 (N_{\nu_x} -N_{\bar\nu_x}) \right]\ ,
\end{eqnarray}
whereas initially the off-diagonal elements vanish. These
expressions represent the energy shift of $\nu_e$ or $\nu_x$
relative to $\nu_s$ caused by matter and neutrino refraction.

\section{Electron fraction evolution}           \label{sec:electronfraction}

The material in a fluid element moving away from the SN core will
experience three stages of nuclear evolution. Near the surface of
the neutron star, typically the material is very hot and essentially
all of the baryons are in the form of free nucleons. As the material
flows away from the neutron star, it cools. When the temperature $T
< 1$~MeV, $\alpha$ particles begin to assemble. As the fluid flows
farther out and cools further, heavier nuclei begin to form. Around
half of the nuclei with masses $A>100$ are supposed to be created by
the r-process, requiring neutron-rich
conditions. In this section, we introduce the $Y_e$ evolution
equation to study whether the impact of sterile neutrinos can help
to produce such an environment.

Having in mind the overall evolution of abundances with radius
and time, namely that close to the neutrino sphere only free
nucleons exist, then alpha-particles begin to form and afterwards
(some) heavy nuclei, the
electron abundance introduced in equation~(\ref{yedef}) can be
expressed as
\begin{equation}
Y_e = X_p + \frac{X_\alpha}{2} + \sum_h \frac{Z_h}{A_h}\,X_h\ .
\end{equation}
Here $X_p$ ($X_\alpha$) is the mass fraction of free protons
(alpha particles) and $Z_h$ and $A_h$ are the charge and mass
number of nuclear species $h$. The summation runs over all
nuclear species $h$
heavier than $\alpha$ particles. However, at the conditions common
to neutrino-heated outflows (in particular in the region where
neutrino interactions have the biggest impact on $Y_e$),
free nucleons and alpha particles
typically account for most of the baryons.

The CC weak interactions alter the electron fraction by converting
neutrons to protons and vice versa.
The electron abundance $Y_e$ in neutrino-heated material
flowing away from the neutron star is set by a competition between
the rates of the following neutrino and antineutrino capture
reactions on free nucleons, assuming that the reactions of neutrinos
on nuclei are negligible,
\begin{eqnarray}
\label{nue1}
\nu_e + n &\rightarrow& p + e^-\ ,\\
\label{anue1}
\bar{\nu}_e + p  &\rightarrow& n + e^+\ ,
\end{eqnarray}
and by the associated reverse processes. Because of slow time variations
of the outflow conditions during the PNS cooling phase, a near
steady-state situation applies and the
$Y_e$ rate-of-change within an
outflowing mass element may be written as~\cite{McLaughlin:1997qi}
\begin{equation}
\label{Yeeq}
\frac{dY_e}{dt} = v(r) \frac{dY_e}{dr} \simeq (\lambda_{\nu_e} + \lambda_{e^+}) Y_n^{\rm f} - (\lambda_{\bar{\nu}_e} + \lambda_{e^-}) Y_p^{\rm f}\ ,
\end{equation}
where $v(r)$ is the velocity of the outflowing mass element and
$Y_n^{\rm f}$ and $Y_p^{\rm f}$ are the abundances of free nucleons.
The forward rates of the neutrino capture processes of
equations~(\ref{nue1}) and (\ref{anue1})
are~\cite{McLaughlin:1997qi}
\begin{eqnarray}
\label{lambdanue}
\lambda_{\nu_e} &\simeq& \frac{L_{\nu_e}}{4 \pi r^2 \langle E_{\nu_e}\rangle}\, \langle \sigma_{\nu_e n}(r)\rangle \ ,
\\
\label{lambdaantinue} \lambda_{\bar{\nu}_e} &\simeq&
\frac{L_{\bar{\nu}_e}}{4 \pi r^2 \langle E_{\bar{\nu}_e} \rangle}\,
\langle \sigma_{\bar{\nu}_e p}(r)\rangle \ .
\end{eqnarray}
The rates for the reverse processes (electron and positron capture
rates on free nucleons) are approximately~\cite{McLaughlin:1997qi}
\begin{eqnarray}
\lambda_{e^-} &\simeq& 1.578 \times 10^{-2}~\mathrm{s}^{-1} \left(\frac{T_e}{m_e}\right)^5 e^{(-1.293 + \mu_{e})/T_e}
\left(1+ \frac{0.646\ \mathrm{MeV}}{T_e} + \frac{0.128\ \mathrm{MeV}^2}{T_e^2}\right)\ ,\\
\lambda_{e^+} &\simeq&1.578 \times 10^{-2}~\mathrm{s}^{-1} \left(\frac{T_e}{m_e}\right)^5 e^{(-0.511 - \mu_{e})/T_e}
\nonumber\\
&&{}\times\left(1+ \frac{1.16\ \mathrm{MeV}}{T_e} + \frac{0.601\ \mathrm{MeV}^2}{T_e^2}+ \frac{0.178\ \mathrm{MeV}^3}{T_e^3}
+ \frac{0.035\ \mathrm{MeV}^4}{T_e^4}\right) \ ,
\end{eqnarray}
where $\mu_e$ is the relativistic electron chemical potential (in MeV).

\begin{figure}[t]
\centering
\epsfig{figure=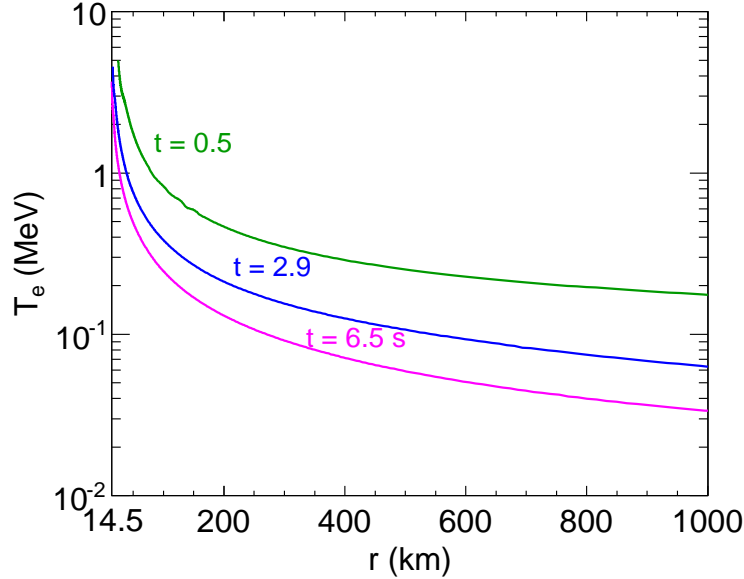, width =11.2cm}
 \caption{Electron temperature as function of radius from the
hydrodynamical simulation of Model Sf $21$~\cite{Huedepohl:2009wh}
for $t = 0.5$, 2.9 and 6.5~s after bounce.
\label{fig1}}
\end{figure}

The electron temperature profile is extracted from the
hydrodynamical simulation for Model Sf $21$ of
\cite{Huedepohl:2009wh} and is shown in figure~\ref{fig1} for the
considered outflow trajectories at different times after core bounce.
The electron chemical potential has been computed by inverting the
equation~\cite{bludman}
\begin{equation}
Y_e = \frac{8 \pi}{3 N_b} T_e^3 \eta (\eta^2 + \pi^2)\ ,
\end{equation}
where $\eta = \mu_e/T$ is the electron degeneracy parameter.

Though the details are more complex, $Y_e$ at small radii is mainly
determined by the $e^-$ and $e^+$ capture rates whereas at larger
radii the reverse neutrino-capture reactions dominate.  Note that in
equation~(\ref{Yeeq}) the nucleons involved in the
$\beta$-reactions are free. Including the corrections due to
nucleons bound in $\alpha$ particles, the free proton and neutron
abundances are
\begin{equation}
Y_p^{\rm f} = Y_e - \frac{X_\alpha}{2}
\quad\hbox{and}\quad
Y_n^{\rm f} = 1 - Y_e - \frac{X_\alpha}{2}\,,
\end{equation}
where $X_\alpha$ is the mass fraction of $\alpha$ particles.
Assuming that nickel-56 is the most abundant among  the heavy
nuclei, we have tested  that the correction to $Y_e$ due to nuclei
heavier than $\alpha$ particles is negligible.
As the electron temperature, the mass fractions of $\alpha$ particles
(and of heavy nuclei for the test) are taken from the
hydrodynamical simulation of Model Sf $21$ of~\cite{Huedepohl:2009wh}.
Because the dominant abundances are free neutrons and protons and
the differences of the $Y_e$ evolution with and without neutrino
oscillations are relatively small, this
is a reasonable approximation, and a consistent treatment
of the composition evolution would yield only minor corrections.

\section{Results}                                       \label{sec:results}
In order to study the impact of sterile neutrinos on $Y_e$ and on
the neutrino fluxes, we discretize the coupled evolution
equations~(\ref{eq:eom1})  in the  energy range 1--60~MeV and
solve them by numerical integration together with
equation~(\ref{Yeeq}). The initial conditions for $Y_e$ from the
hydrodynamical simulation are reported in table~\ref{tab:table2}.
The un-oscillated neutrino fluxes are fixed by the SN model
described in section~2 according to the data given in
table~\ref{tab:table1} for $t=0.5$, 2.9 and 6.5~s.
\begin{table}[t]
\begin{center}
\begin{tabular}{ccc}
\hline
$t$ & $R_\nu$ & $Y_e~(\times 10^{-2})$\\
\hline
0.5 & 25 & 5.47\\
2.9 & 16 & 3.23\\
6.5 & 14.5 & 2.33\\
\hline
\end{tabular}
\vspace{2.mm} \caption{Initial electron abundance at the neutrino
sphere $R_\nu$ (in km) for our three considered post-bounce
times (in s).\label{tab:table2}}
\end{center}
\end{table}
We consider two possible  scenarios: one ``with neutrino
oscillations''  where we include the dynamical feedback on $Y_e$ due
to neutrino oscillations and the other one ``without neutrino
oscillations.''

\subsection[Early cooling phase ($t = 0.5$~s)]{Early cooling phase (\boldmath$t = 0.5$~s)}

\begin{figure}
\centering
\epsfig{figure=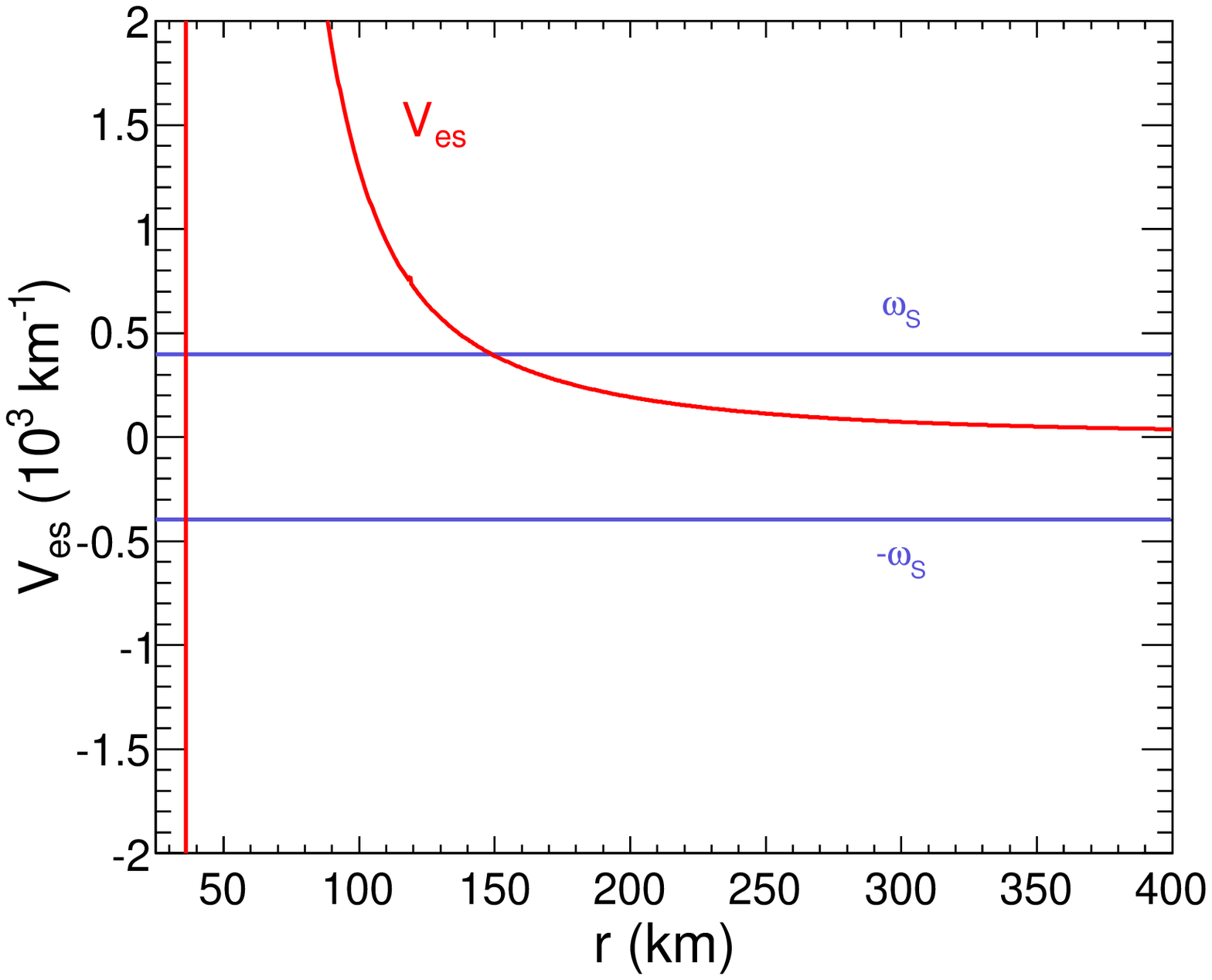, width =11.2cm}
\caption{Refractive energy difference
$V_{es}$ between $\nu_e$ and $\nu_s$ for the
snapshot $t = 0.5$~s.   The horizontal line marks $\pm \omega_{\rm S}$
for a typical neutrino energy of 15 MeV. \label{fig2}}
\end{figure}
\begin{figure}
\centering
\epsfig{figure=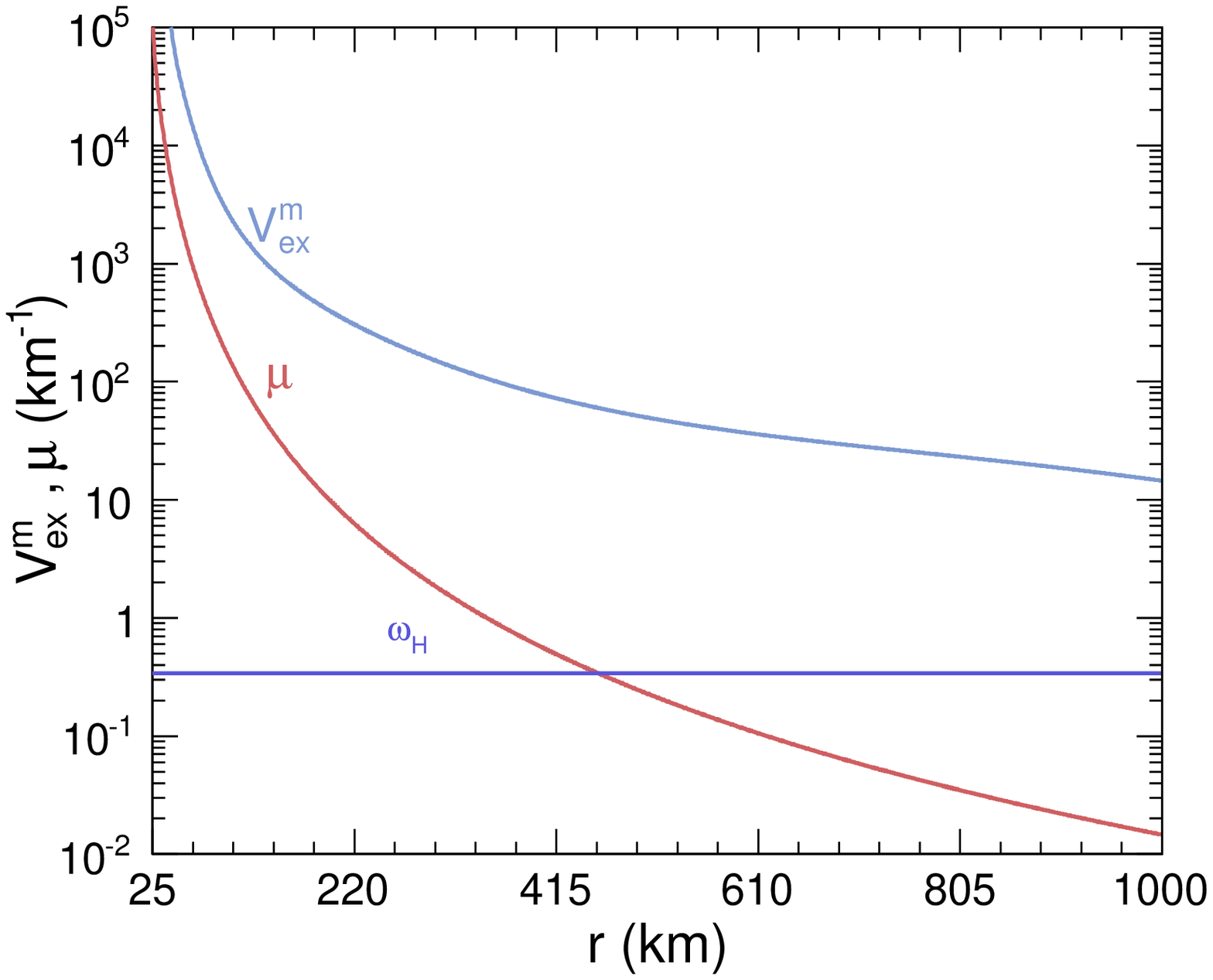, width =11.2cm}
 \caption{Refractive $\nu_e$--$\nu_x$ energy difference caused by matter,
 $V_{ex}^{\rm m}$, and the estimate $\mu$ for the neutrino-neutrino interaction energy
 for the $t=0.5$~s model.  The horizontal line marks $\omega_{\rm H}$
 for a typical neutrino energy of 15 MeV. \label{fig3}}
\end{figure}

\begin{figure}
\centering
\epsfig{figure=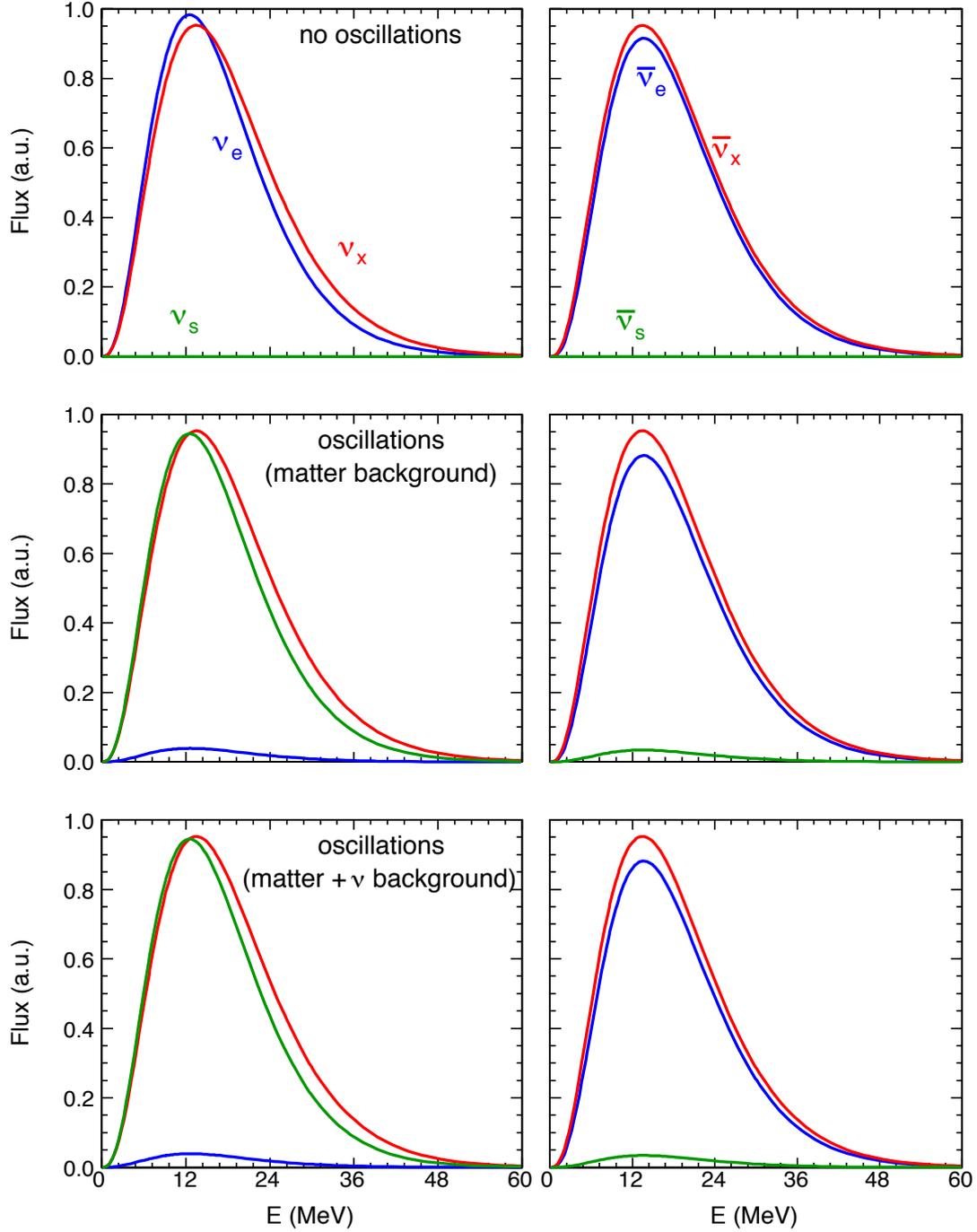, width=16cm}
 \caption{Spectra for neutrinos (left) and antineutrinos (right) at $r
   = 1000$~km in arbitrary units (a.u.) for the 0.5~s model.  Top: No
   oscillations. Middle: Oscillated spectra, including only the matter
   effect. Bottom: $\nu$--$\nu$ interactions are also included, but
   cause no visible difference.\label{fig4}}
\end{figure}

\begin{figure}
\centering
\epsfig{figure=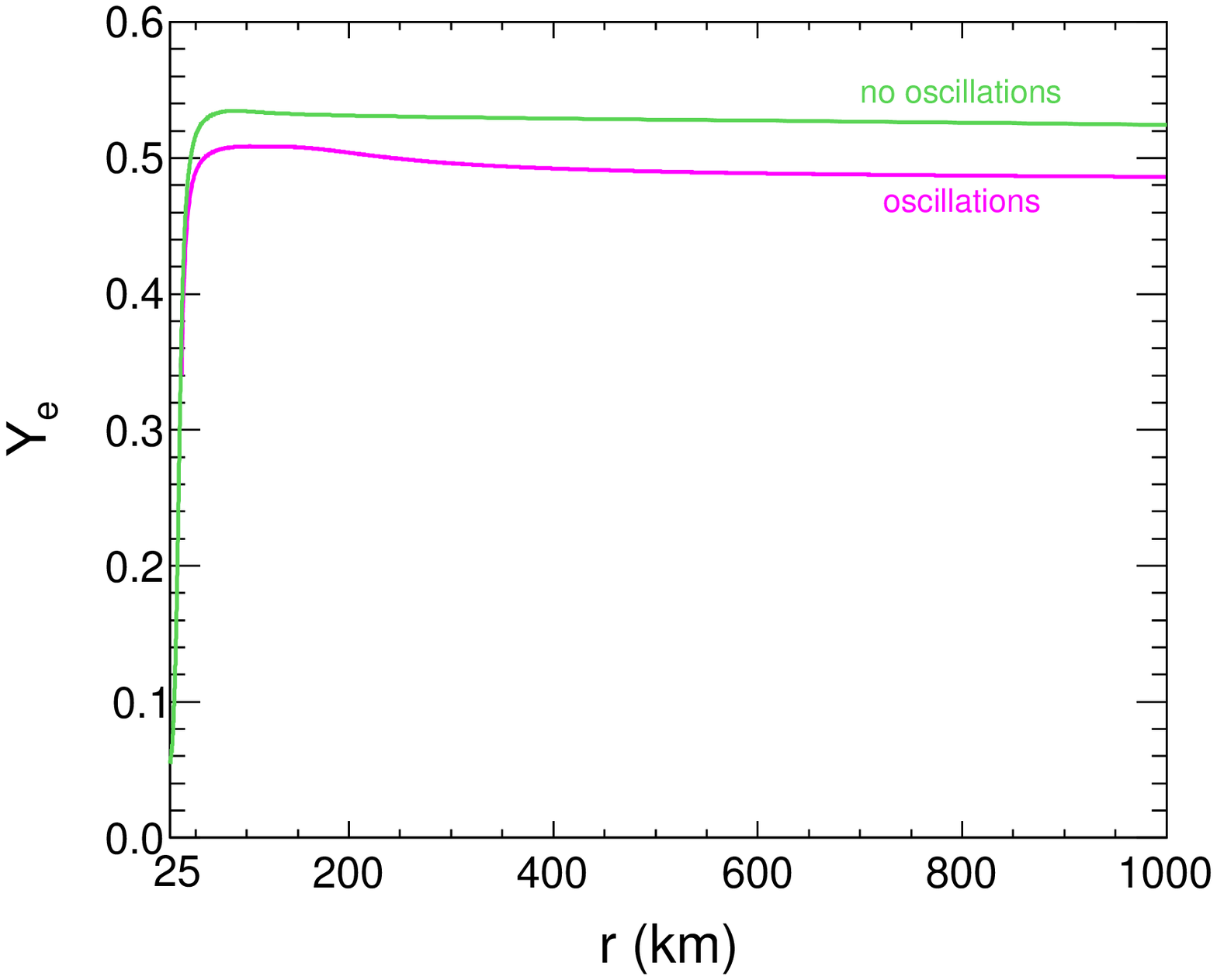, width =11.2cm}
 \caption{Electron abundance for the  $t = 0.5$~s model, with and without
 oscillations.
 \label{fig5}}
\end{figure}

Figure~\ref{fig2} shows the radial profile of the refractive energy
difference $V_{es}=H_{{ee}}^{{\rm m}+\nu\nu}-H_{{ss}}^{{\rm
m}+\nu\nu}=H_{{ee}}^{{\rm m}+\nu\nu}$ between $\nu_e$ and $\nu_s$ as
given in equation~(\ref{lambdaes}). This result already includes a
self-consistent solution for $Y_e$ caused by neutrino oscillations.
The horizontal line marks $\omega_{\rm S}$ for a
representative neutrino energy of 15 MeV.
An MSW resonance arises for both neutrinos and antineutrinos close to
the neutrino-sphere. However, here the effective potential varies
extremely fast because of the sudden rise of $Y_e$, preventing any
significant flavor conversion (extremely non-adiabatic condition).
At larger radii, a second resonance occurs for $r < 300$~km, but
only for neutrinos, causing adiabatic flavor conversion. Therefore,
we expect a larger flavor modification in the neutrino sector than
in the antineutrino one.

Figure~\ref{fig3} shows the radial profile of the
$\nu_e$--$\nu_x$ refractive energy difference caused by matter alone
\begin{equation}
V_{ex}^{\rm m}=\sqrt{2}\,G_{\rm F}\,Y_e N_b\,,
\end{equation}
including the self-consistent solution for $Y_e$. In the active-active sector, collective neutrino oscillations will
occur. Their relevance is not correctly measured by the refractive
energy difference $V_{ex}^{\nu\nu} = \sqrt{2}\,G_{\rm
F}\,(N_{\nu_e}-N_{\bar\nu_e}-N_{\nu_x}+N_{\bar\nu_x})$ because
collective effects are important even if this quantity vanishes due
to the off-diagonal refractive index. One way to express the
possible relevance of collective oscillations is in terms of the
quantity
\begin{equation}
\mu=\sqrt{2}\,G_{\rm F}\,(N_{\nu_e}+N_{\bar\nu_e}+N_{\nu_x}+N_{\bar\nu_x})\,,
\end{equation}
although somewhat different definitions of $\mu$ have been used in
the literature. The potential $\mu$ is plotted in figure~\ref{fig3}
as a function of the radius in order to give a comparison of the
active neutrino abundance with respect to the electron one. For sake
of simplicity, $N_{\nu_\beta}$ are fixed to their initial values
since the correction due to neutrino density variations does not
change the ratio between $\mu$ and
 $V_{ex}^{\rm m}$. The horizontal line marks $\omega_{\rm H}$ for the
neutrino energy 15 MeV. No MSW resonance in the atmospheric sector
is expected and the matter potential is always larger than $\mu$.

In figure~\ref{fig4} we show the spectra for neutrinos (left) and
antineutrino (right). The top panels show the primary spectra at the
neutrino-sphere (no oscillations). Below we show the oscillated
spectra when matter refraction is included, causing an almost
complete MSW swap between $\nu_e$ and $\nu_s$, but hardly any
conversion between $\bar\nu_e$ and $\bar\nu_s$.  Finally in the
bottom panel we show the result after including neutrino-neutrino
interactions in the single-angle approximation, causing no further modification. After the
$\nu_e$--$\nu_s$ MSW conversion, the \hbox{$e$--$x$} difference spectrum is
very asymmetric between neutrinos and antineutrinos, essentially
suppressing collective conversions. Moreover, $F_{\nu_e}(E) <
F_{\nu_x}(E)$ and the same is true for $\bar{\nu}$.
The evolution is therefore inhibited because the system is close
to a stable equilibrium point~\cite{Dasgupta:2009mg,Fogli:2009rd}. 
 Given the initial conditions produced by the active-sterile MSW effect 
for the subsequent collective oscillations, we expect that any multi-angle 
treatment might produce only a smearing of the spectral features without changing
the hierarchy among the fluxes of different flavors~\cite{arXiv:0707.1998,arXiv:0808.0807}. 
In fact the large asymmetry between $\nu_e$ and $\nu_x$ fluxes (and the same for $\bar{\nu}$) 
is responsible for inhibiting any possible flavor decoherence effect~\cite{arXiv:0706.2498}.

Figure~\ref{fig5} shows the effect of oscillations on the $Y_e$
profile. The MSW swap between $\nu_e$ and $\nu_s$ almost
completely removes the original $\nu_e$ flux and pushes the matter
outflow to a more neutron-rich environment, although not far enough
to establish obviously favorable conditions for an r-process.

\subsection[Intermediate cooling phase ($t=2.9$~s)]{Intermediate cooling phase (\boldmath$t=2.9$~s)}

\begin{figure}
\centering
\epsfig{figure=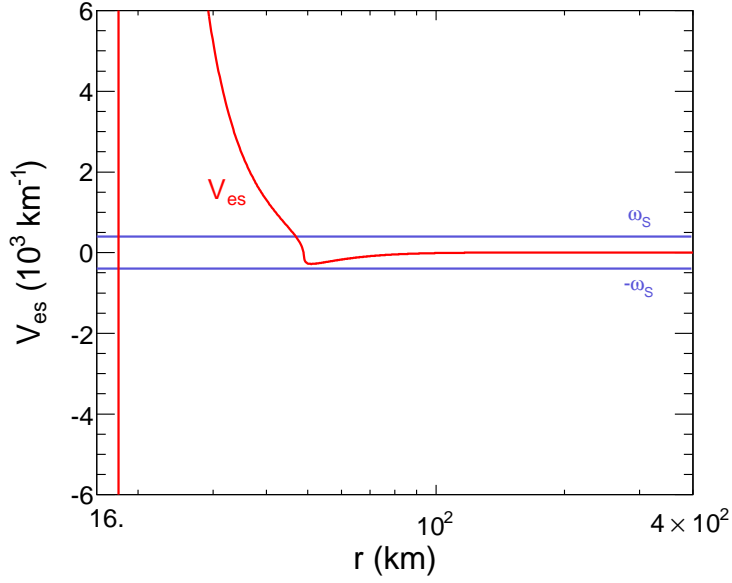, width =11.2cm}
\caption{Refractive energy difference
$V_{es}$ between $\nu_e$ and $\nu_s$ for the
snapshot $t = 2.9$~s.   The horizontal line marks $\pm \omega_{\rm S}$
 for a typical neutrino energy of 15 MeV.
  \label{fig6}}
\end{figure}

\begin{figure}
\centering
\epsfig{figure=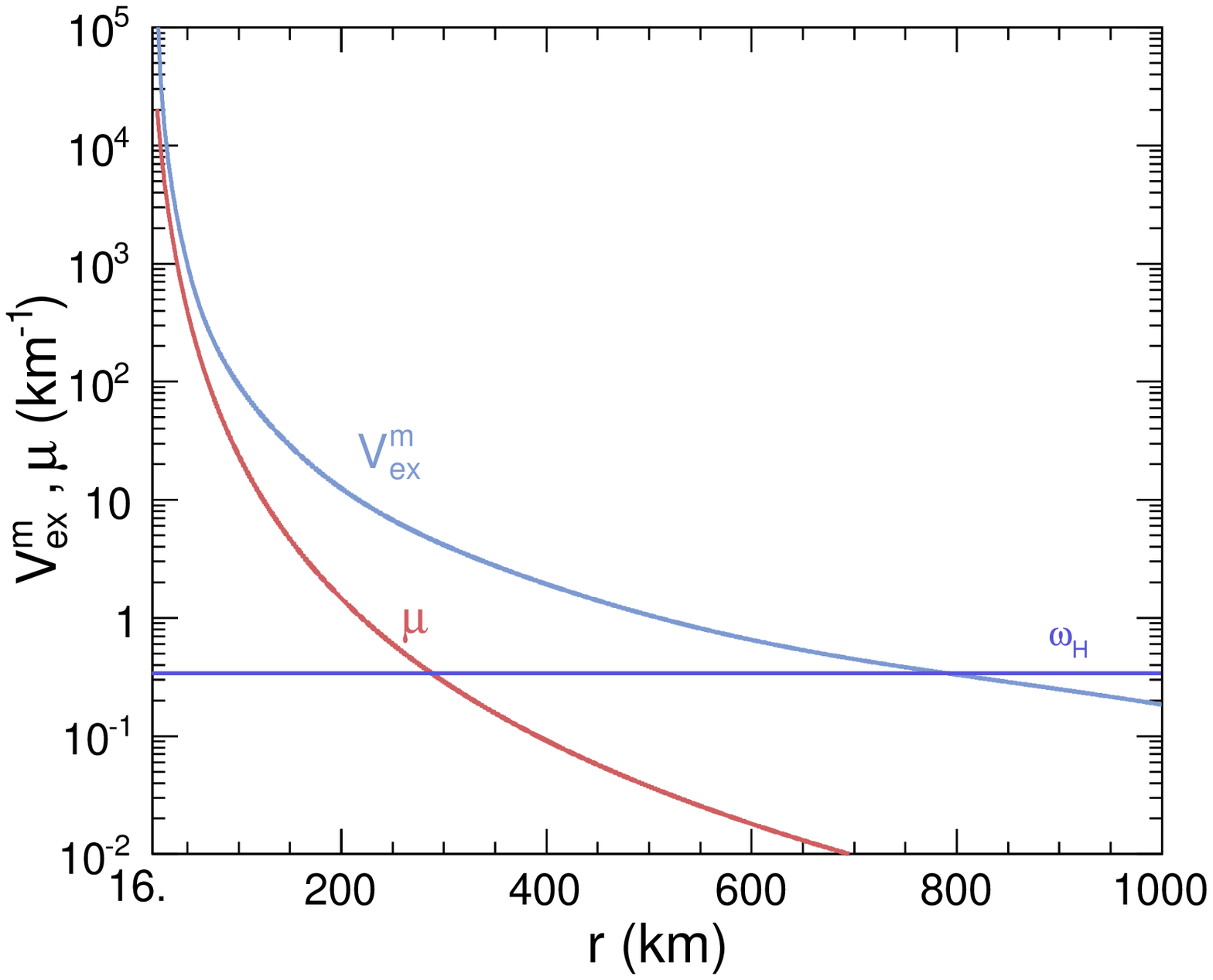, width=11.2cm}
 \caption{Refractive $\nu_e$--$\nu_x$ energy difference caused by matter,
 $V_{ex}^{\rm m}$, and the estimate $\mu$ for the neutrino-neutrino interaction energy
 for the $t=2.9$~s model.  The horizontal line marks $\omega_{\rm H}$
 for a typical neutrino energy of 15 MeV. \label{fig7}}
\end{figure}

\begin{figure}
\centering
\epsfig{figure=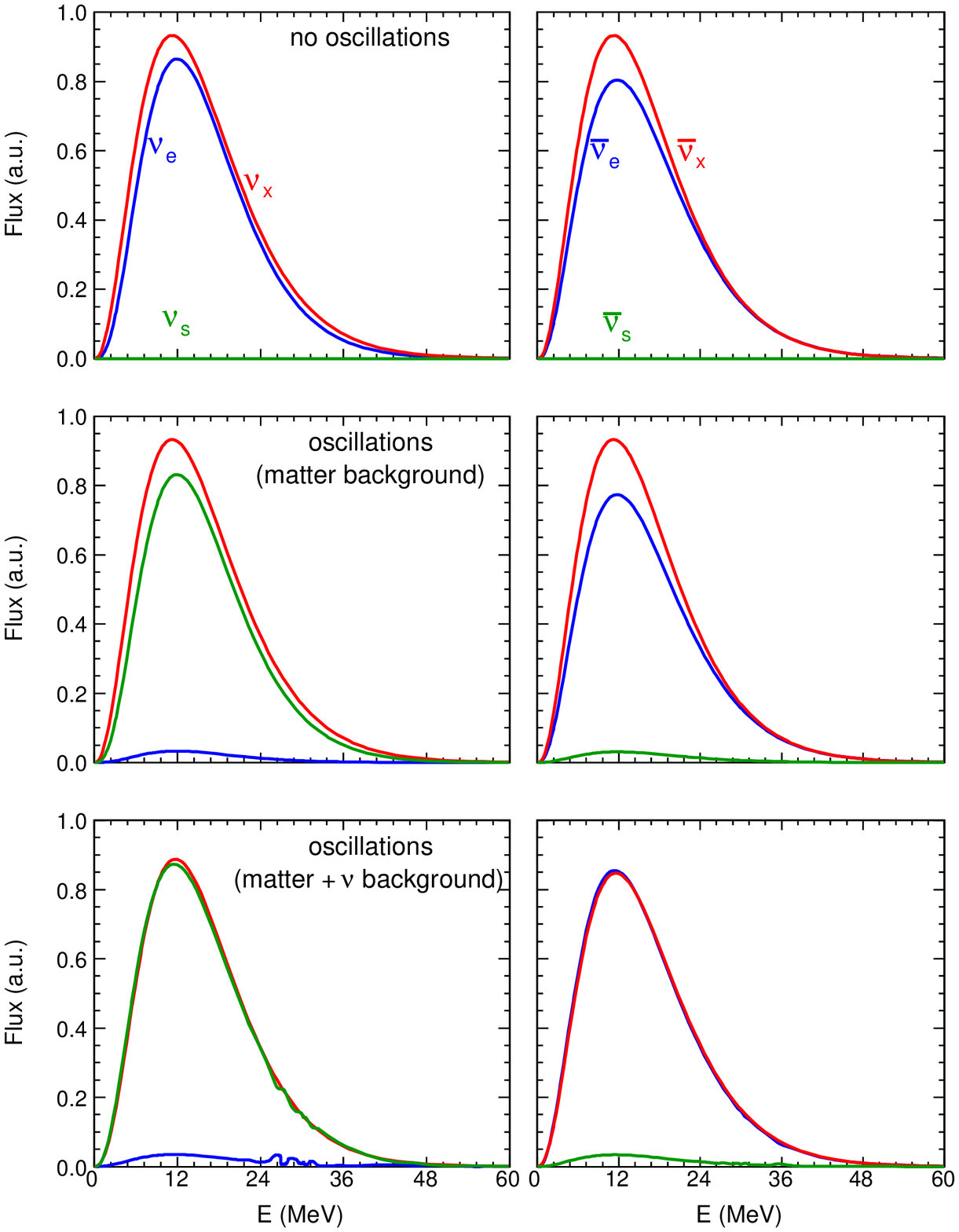, width =16cm}
 \caption{Energy spectra for $t = 2.9$~s as in figure~\ref{fig4}.\label{fig8}}
\end{figure}

\begin{figure}
\centering
\epsfig{figure=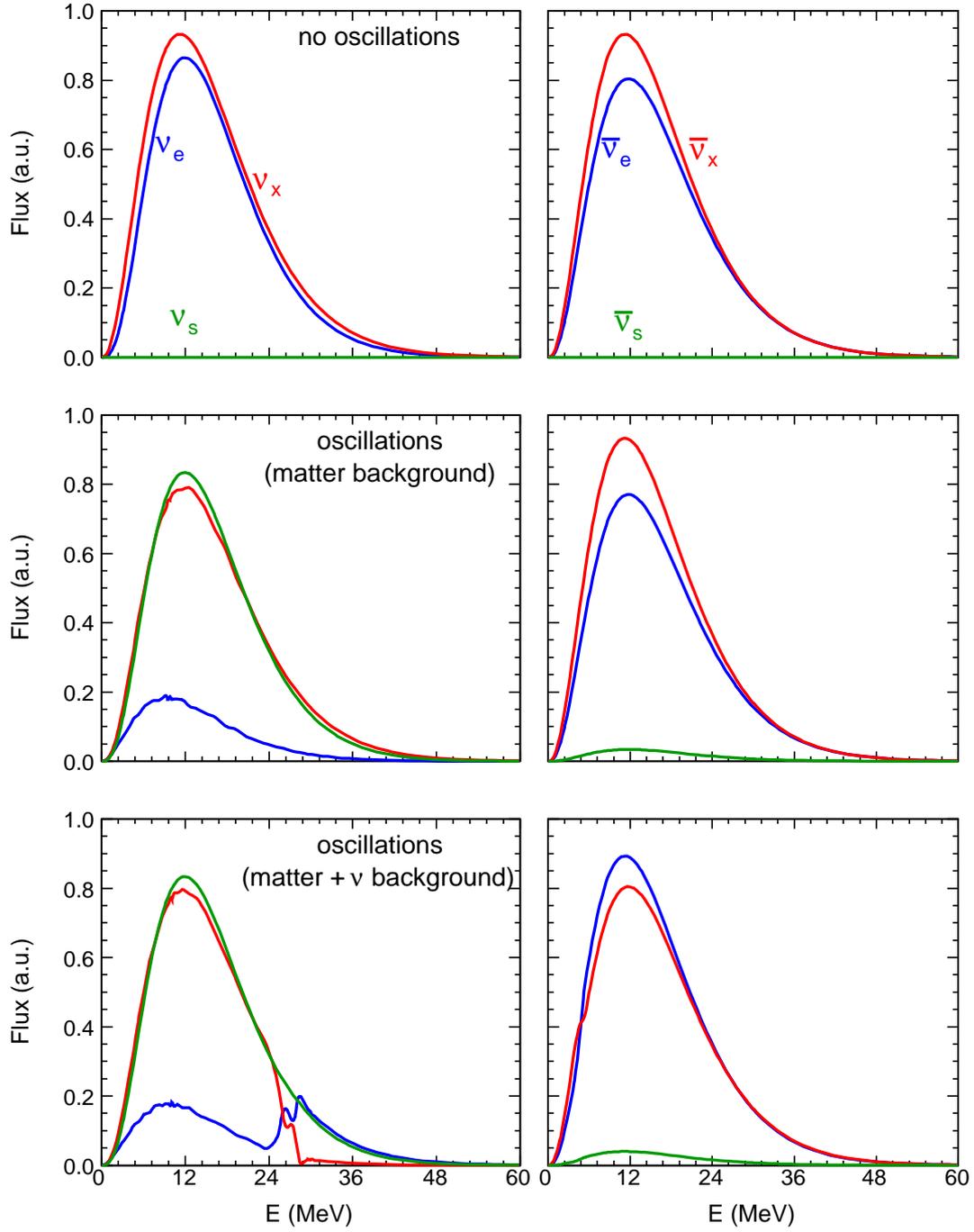, width =16cm}
 \caption{Same as figure~\ref{fig8} with $\sin^2 \Theta_{13} = 10^{-2}$
instead of our usual value $\sin^2 \Theta_{13} = 10^{-4}$. \label{fig10}}
\end{figure}

\begin{figure}
\centering
\epsfig{figure=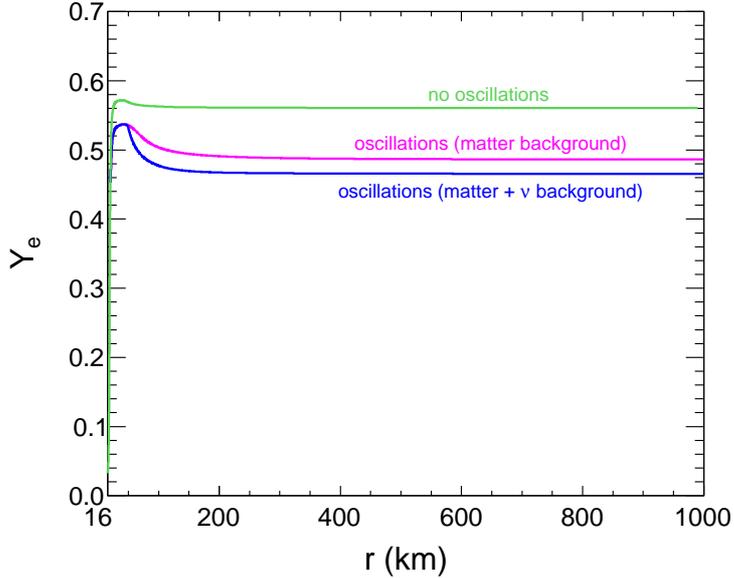, width =11.2cm}
 \caption{Electron abundance for the $t=2.9$~s model for different
 oscillation cases as indicated.
 \label{fig9}}
\end{figure}

We next turn to a snapshot during the intermediate cooling phase at
$t=2.9$~s post bounce. In figures~\ref{fig6}--\ref{fig9} we show the
analogous information as previously for the early cooling phase.
There are several new effects. One is that the refractive difference
between $\nu_e$ and $\nu_{\rm s}$ quickly drops in an almost
step-like feature, caused by $\nu_e$--$\nu_{s}$ MSW conversions
(figure~\ref{fig6}). Another is that the $\nu_e$--$\nu_x$ refractive
energy difference caused by matter is now much smaller, allowing for
an MSW effect between the two active flavors in the neutrino sector
for the chosen hierarchy (figure~\ref{fig7}). The neutrino
background is responsible for increasing the  $\bar{\nu}_e$ flux
with respect to the case with only matter and for  averaging out the
$\bar{\nu}_x$ and $\bar{\nu}_e$ fluxes.

In figure~\ref{fig9} the electron abundance is plotted as a function
of the radius. The $Y_e$ profile is lowered compared to the
no-oscillation case. In particular, when $\nu\nu$ refraction is
included, the asymptotic $Y_e$ value, due to the increase of
$\bar{\nu}_e$ flux, is more significantly shifted below 0.5 than
at the earlier time of 0.5~s after bounce.

For comparison, in figure~\ref{fig10} the neutrino and antineutrino
fluxes are shown for larger $\sin^2 \Theta_{13}$. In this case the
MSW $\nu_e$--$\nu_x$ conversion is adiabatic and as a result, the
matter-only oscillations not only lead to an almost complete swap of
$\nu_{s}$ to $\nu_e$, but in addition a partial $\nu_e$--$\nu_x$
swap. Including neutrino-neutrino refraction, since collective
effects and MSW resonances are occurring all in the same spatial
range, a spectral swap with two splits emerges, one in the neutrino
sector at about 26~MeV, the other in the anti-neutrino sector at
about 6~MeV. However, the increase of the $\nu_e$ flux at high
energies counter-balances the increase of the $\bar{\nu}_e$ flux in
such a way that the resultant $Y_e$ does slightly change but without
modifying our conclusions.

\clearpage


\clearpage

\subsection[Late cooling phase ($t=6.5$~s)]{Late cooling phase (\boldmath$t=6.5$~s)}

\begin{figure}
\centering
\epsfig{figure=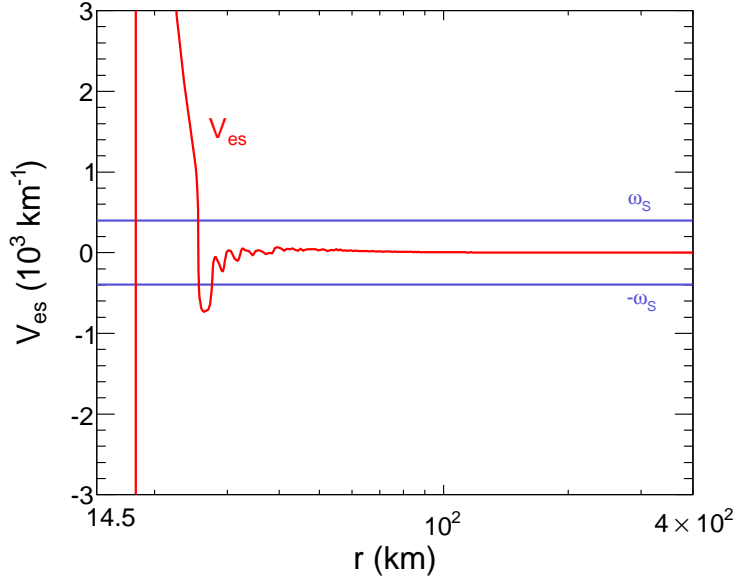, width =11.2cm}
 \caption{Refractive energy difference
$V_{es}$ between $\nu_e$ and $\nu_s$ for the
snapshot $t = 6.5$~s.   The horizontal line marks $\pm \omega_{\rm S}$
 for a typical neutrino energy of 15 MeV. \label{fig11}}
\end{figure}

\begin{figure}
\centering
\epsfig{figure=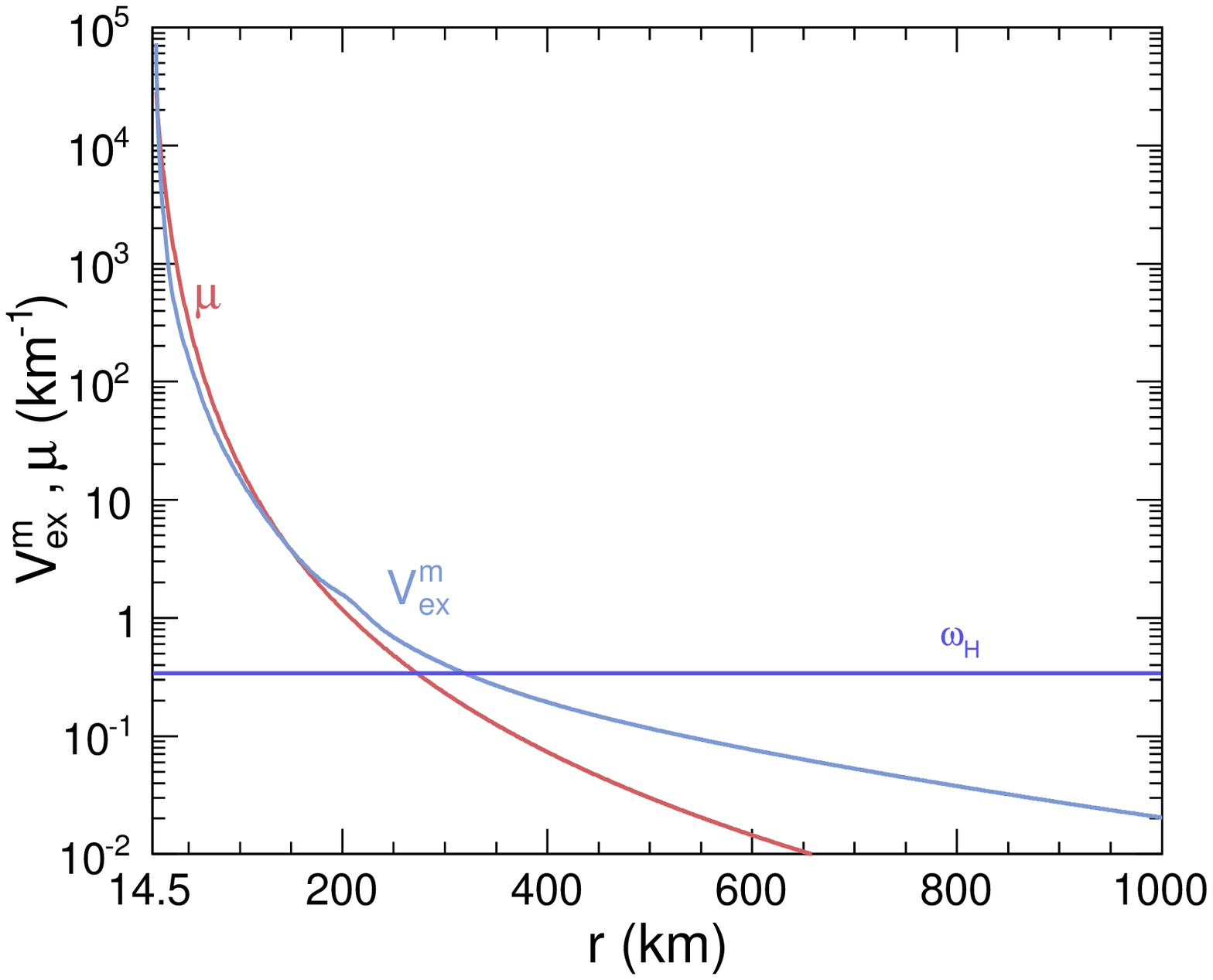, width =11.2cm}
 \caption{Refractive $\nu_e$--$\nu_x$ energy difference caused by matter,
 $V_{ex}^{\rm m}$, and the estimate $\mu$ for the neutrino-neutrino interaction energy
 for the $t=6.5$~s model.  The horizontal line marks $\omega_{\rm H}$
 for a typical neutrino energy of 15 MeV.  \label{fig12}}
\end{figure}

\begin{figure}
\centering
\epsfig{figure=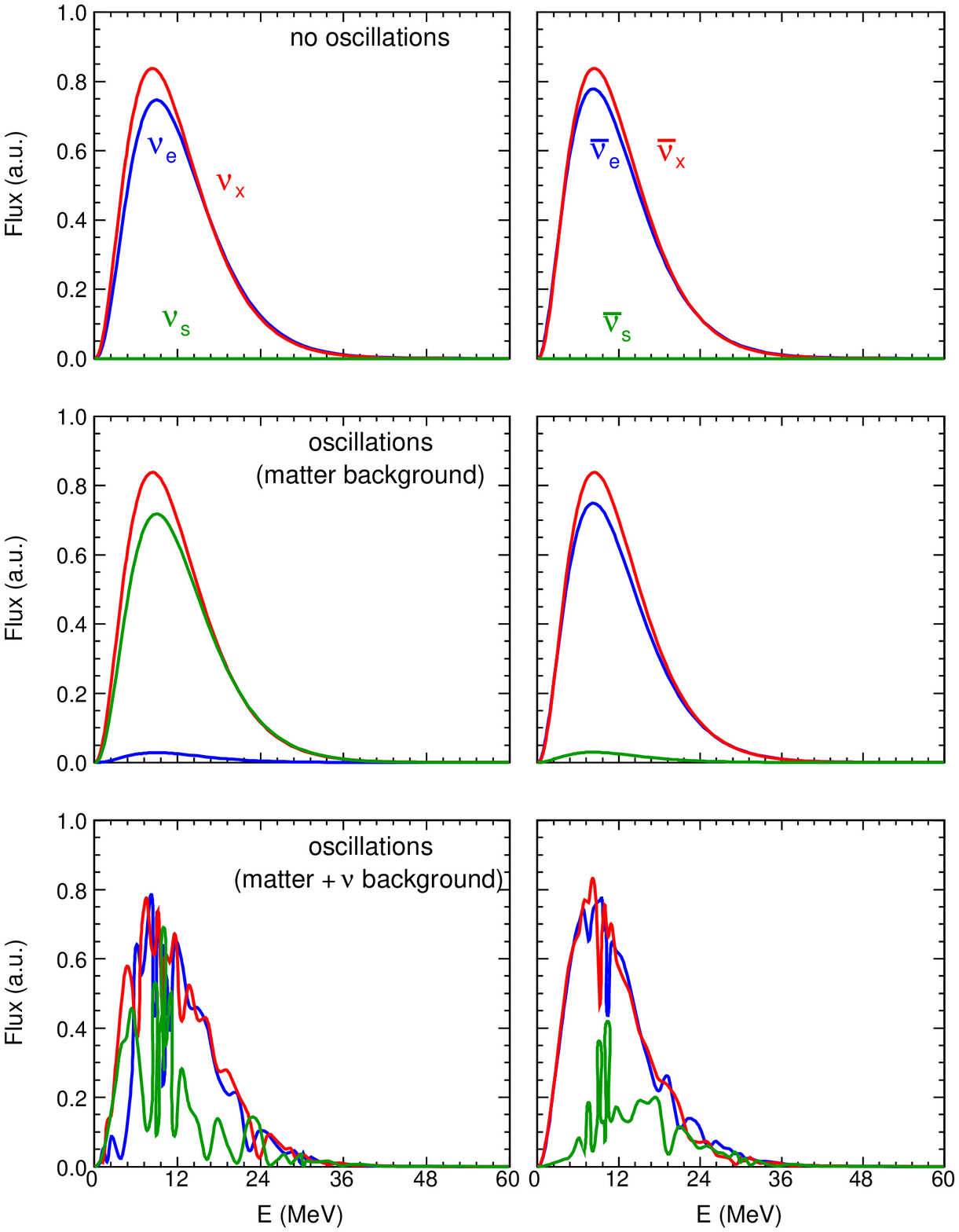, width =16cm}
 \caption{Energy spectra for $t = 6.5$~s as in figure~\ref{fig4}.\label{fig13}}
\end{figure}

\begin{figure}[t]
\centering
\epsfig{figure=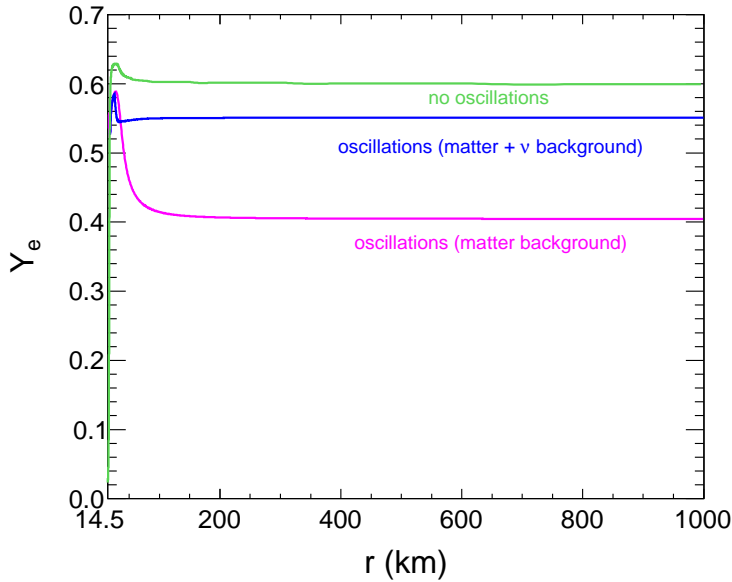, width =11.2cm}
 \caption{Electron abundance evolution for the $t=6.5$~s model for
 different oscillation cases as indicated.
 \label{fig14}}
\end{figure}

Finally, we consider the late cooling phase for a snapshot at $t=6.5$~s
post bounce. In figures~\ref{fig11}--\ref{fig14} we show the
analogous information as in the previous cases. The active-sterile
MSW effect once more leads to an almost complete $\nu_e$--$\nu_{s}$
swap, if we include only the ordinary matter effect. However, once we
include $\nu\nu$ refraction, the results change quite dramatically.
The active-sterile energy difference again drops quickly at a
critical radius because neutrinos contribute significantly to the
matter effects and the MSW conversion shifts the total matter effect
to zero and then has oscillatory features, apparently causing
parametric resonance effects in subsequent neutrino oscillations.
Flavor conversions differ for each energy mode, inducing wiggles in
the energy spectra shown on the bottom panels of figure~\ref{fig13}.
As a result, the oscillated  spectra are mixed up and, in
particular, the $\nu_e$ and $\nu_x$ spectra are almost coincident.
It is  $\nu\nu$ interactions that are responsible for repopulating
the $\nu_e$ spectrum and, consequently increase the $Y_e$ value with
respect to the case with only matter background as visible in
figure~\ref{fig14}. The asymptotic $Y_e$ value drops far below
0.5 when only the matter effect is considered, but including $\nu\nu$
interactions it is pushed back up above 0.5, although in both cases
$Y_e$ is lowered relative to the no-oscillation case.


\clearpage

\section{Conclusions}           \label{sec:conclusions}

Motivated by recent hints for the existence of sterile neutrinos,
and in particular the antineutrino reactor anomaly, we have studied
flavor oscillations within a three-flavor $(\nu_e,\nu_x,\nu_{s})$
scheme in the context of an electron-capture SN model, focussing on
the neutrino-cooling phase of the nascent neutron star. We have
included $\nu_e$-$\nu_s$ mixing with parameters suggested by the
reactor anomaly and active-active mixing representing 1--3
oscillations driven by the atmospheric mass difference and a small
$\Theta_{13}$ angle. Our main goal was the determination of neutrino
flux and spectral changes and their impact on the evolution and
asymptotic value of $Y_e$ in the neutrino-driven wind ejecta. Their
modification is relevant for nucleosynthesis in SN outflows. We have
studied three snapshots that are representative for the early,
intermediate, and late cooling stages.

Even in our simplified neutrino mixing scheme, the results depend
sensitively and in complicated ways on the detailed matter profile
and neutrino fluxes and spectra. In the early phase after the onset
of the explosion, oscillations are driven almost entirely by the
ordinary matter effect and lead to a simple $\nu_e$--$\nu_{s}$ MSW
conversion. In the latter cases, neutrinos contribute significantly
to the refractive energy shifts. One result is that the overall
$\nu_e$--$\nu_{s}$ energy difference drops to zero when some of the
MSW conversion is complete and in this way oscillations feed back on
themselves. The switch-off of the matter effect apparently can lead
to parametric resonance effects and a repopulation of $\nu_e$ from
$\nu_x$.

In all cases, the asymptotic $Y_e$ value is lowered compared to the
non-oscillation case, but it sensitively depends on the cooling
phase how large this effect turns out to be. The inclusion of
active-active collective oscillations and MSW conversions in
addition to active-sterile mixing strongly modifies the outcome.
These general trends are not severely altered by larger values of
$\Theta_{13}$.

Accordingly, neutrino conversion to a sterile flavor as well as
oscillations and collective transformations of active flavors
influence the radial variation and time-dependent asymptotic value
of $Y_e$ in the neutrino-driven wind in complicated ways. The
feedback of active-sterile oscillations on the refractive effect
causes intriguing nonlinear modifications of the naive oscillation
picture and active-active oscillations can in addition play an
important role in determining the neutron-to-proton ratio of SN
ejecta.

The chemical composition of the matter outflow can thus
be strongly affected by neutrino oscillations. In our model of the
neutrino cooling of the proto-neutron star born in an
electron-capture SN, the corresponding changes do not lead to a
large neutron excess. Therefore it appears unlikely that in the studied
cases viable r-process conditions could be produced by flavor
oscillations, although the formation of heavy elements must be 
expected to change.

All numerical studies of neutrino oscillations with
  neutrino-neutrino refraction use simplified assumptions, in our case
  the ``single angle approximation,'' but even more sophisticated
  ``multi-angle studies'' assume axial symmetry around the radial
  direction and therefore are not fully general.  Even if we suspect that 
  our conclusions are not affected by the $\nu$--$\nu$ interaction's angular dependence, a 
  careful analysis on this effect remains certainly to be investigated.

More important may be the calculation of a denser grid of snapshots
for more solid conclusions on the nucleosynthetic implications and as
the basis for detailed studies of element formation.  Moreover,
because the neutrino emission properties and the neutrino-driven wind
conditions depend sensitively on the mass of the proto-neutron
star~\cite{Qian:1996}, our results for the 1.36\,$M_\odot$ (baryonic
mass) remnant of an electron-capture SN may be applicable only to SNe
with similar compact objects.  The investigation of a broader range of
progenitor models, in particular also of iron-core SNe with more
massive proto-neutron stars, is therefore desirable to identify
possible cases where favorable conditions for the r-process may be
produced by flavor oscillations involving sterile neutrinos.

If sterile neutrinos with parameters suggested by the presently
discussed antineutrino reactor anomaly are verified in future
experiments, their existence cannot be ignored in nucleosynthesis
studies of the SN environment. Sterile neutrinos must be expected to
have important consequences for the possibility of a $\nu$p-process
in SN outflows and might even be relevant for the question whether
SN explosions can be sources of r-process elements. The experimental
neutrino mixing parameters will be a crucial input information for
theoretical investigations of these problems.

\section*{Acknowledgments} 

We acknowledge partial support from the Deutsche
Forschungsgemeinschaft by grants\ TR~7, TR~27 and EXC~153.
I.T.\ acknowledges support from the Alexander von Humboldt
Foundation.


\end{document}